\documentstyle[12pt,aaspp4]{article}

\begin{document}

\title{The Future of Extragalactic Research}
\author{Sidney van den Bergh}
\affil{National Research Council of Canada \\ 
Dominion Astrophysical Observatory \\
5071 West Saanich Road \\
Victoria, British Columbia, V8X 4M6 \\
Canada }

\bigskip
\bigskip
\hspace{2.5in}
\underline{There are fools who pretend to predict the future}

\hspace{4.9in} Erasmus (1510)
\bigskip

\begin{abstract}
	It is argued that the astronomy of the twenty-first 
century will be dominated by computer-based manipulation 
of huge homogeneous surveys of various types of astronomical 
objects.  Furthermore combination of all observations with 
large telescopes into a single database will allow data 
mining on an unprecedented scale.
\end{abstract}

\section{Introduction}
	Predictions regarding the future directions of research 
in science have a very poor track record.  The main reason for 
this is, no doubt, that the most important developments in science 
are also the most unexpected ones.  It follows that the great 
discoveries in astronomy cannot be predicted by linear 
extrapolation from past trends.  Furthermore, serendipitous 
discovery plays a major role in the advancement of our science.  
Nevertheless, it is perhaps safe to let one's thinking about the future be guided by George Ellery Hale's notion that progress in observational astronomy always requires ``Light! More light!'' (Woodbury, 1944). During recent decades the greatest advance has come from the replacement of photographic plates by charge coupled detectors.  Such CCDs increase the efficiency of photon detection by two orders of magnitude.  Furthermore they produce data in a format that can be easily manipulated by digital computers.  An additional gain by a factor of $\sim 1 \times 10^2$ in global light gathering power has resulted by going from Hale's single 5-m reflector on Palomar Mountain to a world-wide arsenal of a dozen or so 5-m to 10-m telescopes.  Since the efficiency of CCD detectors is already close to 100\% little additional gain can be expected from increases in detector speed.  Although significant increases in telescope aperture are technically possible it seems doubtful whether Society would be willing to support the expenditures required to increase the size of telescopes by two orders of magnitude.  The only remaining way to increase the number of photons available to astronomers would appear to be by increasing detector size, and hence the area of the sky, that can be imaged at any given time.  Such wide-field detectors will make it possible to undertake enormous homogeneous surveys of various classes of astronomical objects.  One might also dream of detectors that could determine the wavelength of each incident photon.  This would allow one to avoid the inefficiencies that are inherent in photometry through intermediate-band filters and in (low dispersion) spectroscopy. 

	On visiting the offices of the Mount Wilson and Palomar Observatories in the nineteen-sixties I was shown one or more plates of the Crab Nebula by four individuals.  Each of these astronomers jealously guarded such plates in his own private files!  Clearly such a situation, where large telescope data do not (after a short proprietary period) become part of the common heritage of mankind, is highly inefficient.  In future all large telescope observations will be available in large publicly available digital data bases where they can be mined, or combined, at will.  This will make it possible for all photons collected globally over many years to contribute to the progress of astronomy.

\section{Galaxies}
	A few years ago a funding agency asked me to help evaluate the proposal for the \underline{Sloan Digital Sky Survey}.  On reading this document I was suddenly struck by the realization that the acquisition and manipulation of such enormous data bases would become central to the astronomy of the twenty-first century.  In this connection one is reminded of Joseph Stalin's dictum that quantity [of tanks] has a quality all its own.  By the same token a computer chip with millions of diodes on its surface has become qualitatively different from a device that contains a few vacuum tubes.  Similarly the human brain, which contains millions of neurons, allows us to explore ``dimensions'' that are not available to a small-brained mouse.

	The availability of homogeneous surveys of millions of galaxies should allow us to gain deep new insights into many aspects of galaxy evolution.  The accuracy of automatic classifications of individual galaxy images by ``neural networks'' is greatly inferior to that of craftsmen like G. de Vaucouleurs, A.R. Sandage, and W.W. Morgan.  However, such mass-produced computer classifications will allow one to investigate the differences between the distributions of millions of galaxies of differing morphological type over the sky. Furthermore enormous homogeneous data sets on galaxies at different redshifts will make it possible to study both the evolutionary histories of different types of galaxies, and the changes in merger rates between galaxies, over a significant fraction of the lifetime of the Universe.  Finally very large homogeneous samples of galaxies may enable one to identify very rare (or unusual) objects that might be worthy of more detailed study.

It is perhaps ironic that such a future style of astronomical observation, which is based on large data samples, will be more similar to that employed by Harlow Shapley and his collaborators at Harvard in the nineteen-thirties, than it is to the modern Palomar/KPNO style of observing, which tends to involve more  intensive study of individual objects and small samples.

\section{Quasars}
	Astronomers appear to be particularly attracted to violent events, such as supernova explosions and outbursts in active galactic nuclei.  Supernovae are of enormous interest because such events (1) liberate vast amounts of energy, (2) produce a large fraction of the heavy elements that are required to sustain life, (3) as possible calibrators of the extragalactic distance scale, and (4) because they form black holes.  It appears probable that individual supernovae of Type II might be detectable at greater redshifts than those at which the first generation of metal-poor galaxies are visible.  Large digital sky surveys might therefore be used to generate samples of ancient supernovae at the edge of the observable Universe.

	Large, deep, and homogeneous surveys should also allow one to study rare quasars with unusually large redshifts, that were formed when the Universe was still quite young.  Furthermore such surveys would probably turn up many unusual active nuclei that would warrant more detailed study with large narrow-field telescopes.  One should, however, heed the warning by Erasmus (1510) that ``We fools have a particular trick of liking best whatever comes to us from farthest away.''  One should therefore never forget that much can also be learned from detailed study of nearby galaxies, such as the members of the Local Group, and from large homogeneous surveys that can be used for systematic discovery of the oldest and most metal-poor objects in the Milky Way.

\section{The Future}
	One should always expect the unexpected.  The 100-inch Hooker  telescope on Mt. Wilson was mainly built to provide the light-gathering power  needed for high-dispersion spectroscopy of stars, but its main claim to fame is that  it allowed Hubble (1929) to discover the expansion of the Universe.  By the same  token the Palomar 5-m reflector was mainly built to establish the extragalactic  distance scale, but it will probably go down in history as the telescope that  discovered quasars.

	Harlow Shapley once remarked that one should hang ribbons around telescopes, rather than medals around the necks of famous astronomers.  Recent experience with the Keck telescope and Hubble Space Telescope tends to support this view.  The development of more powerful telescopes has clearly been the dominant driving force behind many of the most spectacular advances of modern astronomy.  Nevertheless it might turn out that larger detectors, and faster computers to analyze their output, will be the most important drivers of twenty-first century astronomy.  This confirms the view of baseball great Yogi Berra that ``The future isn't what it used to be''.

\end{document}